\documentclass{svjour3}
\usepackage[utf8]{inputenc}
\usepackage{graphicx}
\usepackage{natbib}
\usepackage[final]{changes}
\usepackage[colorlinks,citecolor=blue]{hyperref}
\usepackage{ifthen}  

\def\glc{{\normalfont \scshape Gal\-ac\-ti\-cus}}

\newcounter{CDMDone}
\def\CDM{\ifthenelse{\equal{\arabic{CDMDone}}{0}}{cold dark matter (CDM)\setcounter{CDMDone}{1}}{CDM}}

\newcounter{CGMDone}
\def\CGM{\ifthenelse{\equal{\arabic{CGMDone}}{0}}{circumgalactic medium (CGM)\setcounter{CGMDone}{1}}{CGM}}

\title{Achieving Convergence in Galaxy Formation Models by Augmenting N-body Merger Trees}
\author{Andrew J. Benson \and Chris Cannella \and Shaun Cole}
\institute{Andrew J. Benson \at Carnegie Observatories, 813 Santa Barbara Street, Pasadena, CA 91101, USA \email{abenson@carnegiescience.edu} \and Chris Cannella \at California Institute of Technology, 1200 E. California Blvd., Pasadena, CA 91125, USA \and Shaun Cole \at Institute for Computational Cosmology, Department of Physics, Durham University, South Road, Durham DH1 3LE, UK}

\begin{document}

\twocolumn

\maketitle

\begin{abstract}
Accurate modeling of galaxy formation in a hierarchical, cold dark matter universe requires the use of sufficiently high-resolution merger trees to obtain convergence in the predicted properties of galaxies. When semi-analytic galaxy formation models are applied to cosmological N-body simulation merger trees, it is often the case that those trees have insufficient resolution to give converged galaxy properties. We demonstrate a method to augment the resolution of N-body merger trees by grafting in branches of Monte Carlo merger trees with higher resolution, but which are consistent with the pre-existing branches in the N-body tree. We show that this approach leads to converged galaxy properties.

\keywords{keywords go here}
\end{abstract}

\section{Introduction}

A commonly used method for generating catalogs of galaxies from theoretical models is to apply semi-analytic galaxy formation models \citep{baugh_primer_2006} to merger trees of dark matter halos extracted from N-body simulations \citep{kauffmann_clustering_1999,helly_galaxy_2003}. This approach has been used to both study the physics of galaxy formation \citep{henriques_monte_2009,bower_parameter_2010,lu_bayesian_2011,mutch_constraining_2013,benson_building_2014,ruiz_calibration_2015} and to generate mock catalogs for use in determining survey sizes and accuracy \citep{lemson_halo_2006,bernyk_theoretical_2016} and comparing models and observations on an equal footing \citep[e.g.][]{farrow_galaxy_2015}. 

A limiting factor in this approach is the resolution of the N-body simulation. A dark matter halo must be resolved with of order 100 particles to have robustly determined properties \citep{velliscig_alignment_2015}. In a modern cosmological simulation (e.g. the MultiDark Planck simulation; \citealt{klypin_multidark_2016}) this implies a minimum resolved halo mass of $\sim 2 \times 10^{11}{\rm M}_\odot$. While this may be sufficiently low to host galaxies of the masses of interest for a given application this does not mean that the physical properties of such galaxies are converged. Galaxy formation in the \CDM\ paradigm is inherently hierarchical, meaning that the properties of a galaxy in a given halo may depend on those of galaxies which form in lower mass progenitor halos. If those halos are not resolved, the properties of the primary galaxy may not be converged even though its halo is resolved. This can happen not only because galaxies in unresolved progenitor halos are not available to merge with the primary galaxy, but also because those galaxies will lock up baryonic material into stars (making it unavailable to the primary galaxy), and produce metals which may contaminate the \CGM\ of the primary galaxy (affecting cooling rates etc.).

For example, \cite{bower_breaking_2006} had to introduce a sub-resolution metal enrichment prescription into their semi-analytic model (to account for metals which would have been produced by unresolved galaxies in a higher resolution simulation) when applying it to merger trees from the Millennium simulation \citep{springel_simulations_2005} in order to achieve the same results as when run on higher resolution merger trees to which they had calibrated their model. 

Rather than introducing sub-resolution models to counteract the lack of resolution, we propose a method which augments N-body merger trees with high-resolution branches generated using a Monte Carlo approach. The branches generated are consistent with the resolved structure of the N-body tree, but can include much higher resolution structure. While these branches lack positional information about their halos, this is generally not necessary to perform the galaxy formation calculations.

The remainder of this paper is organized as follows. In \S\ref{sec:Method} we describe our method for augmenting the resolution of N-body merger trees, and in \S\ref{sec:Results} we demonstrate how this approach leads to convergence in galaxy properties (specifically stellar masses are explored in this work, although our approach works for all galaxy properties). Finally, in \S\ref{sec:Conclusions} we give our conclusions.

\section{Method}\label{sec:Method}

Given a merger tree extracted from an N-body simulation, our goal is to augment the resolution of the tree by grafting on higher resolution branches which match the pre-existing tree structure to some given precision. To generate those branches we will use the \cite{parkinson_generating_2008} algorithm, which has been tuned to reproduce the statistical properties of N-body merger trees. Hereafter we refer to trees generated using this algorithm as PCH trees.

\begin{figure*}
 \begin{tabular}{cc}
   \includegraphics[width=75mm,bb=0 300 612 792]{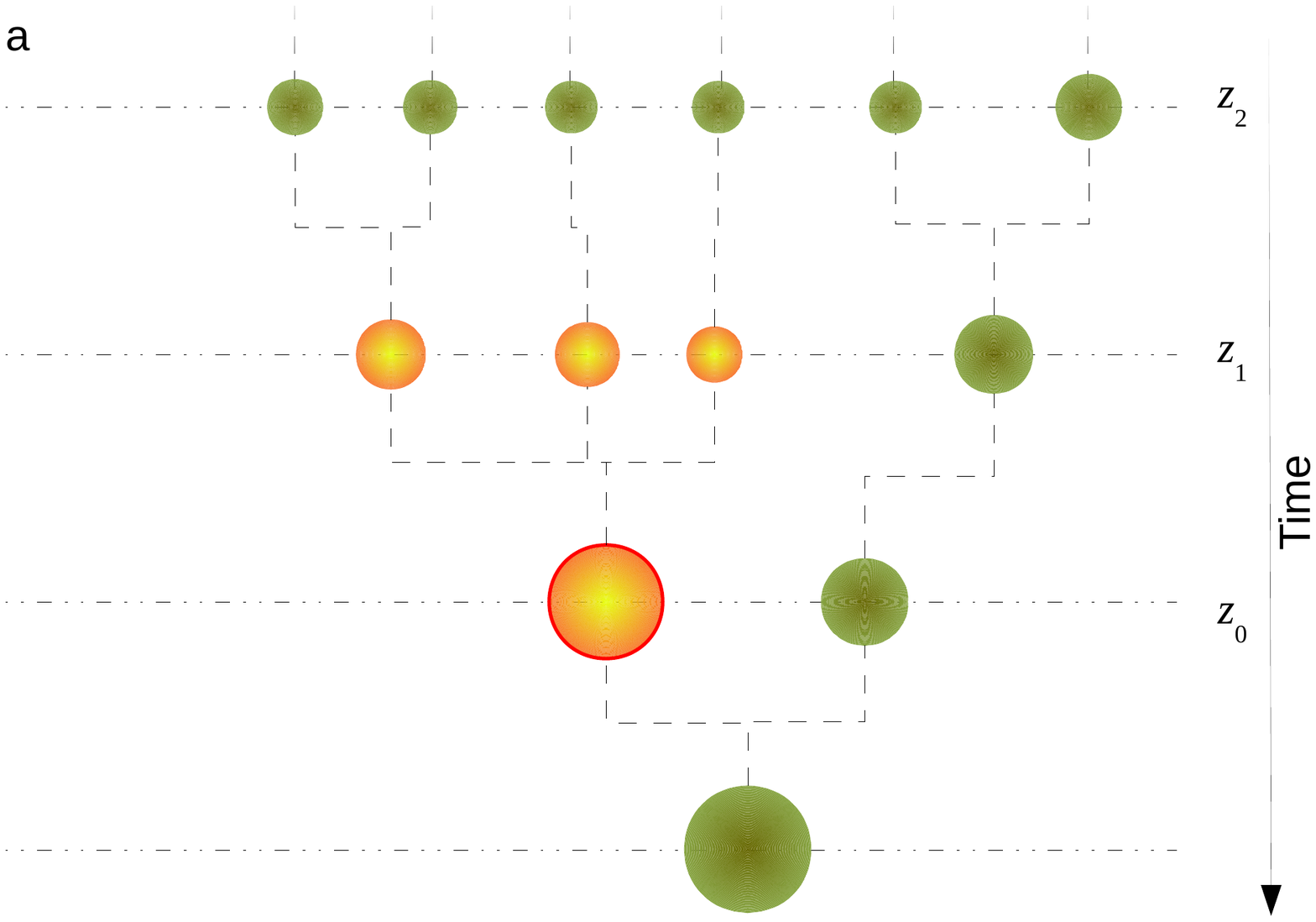} &
   \includegraphics[width=75mm,bb=0 300 612 792]{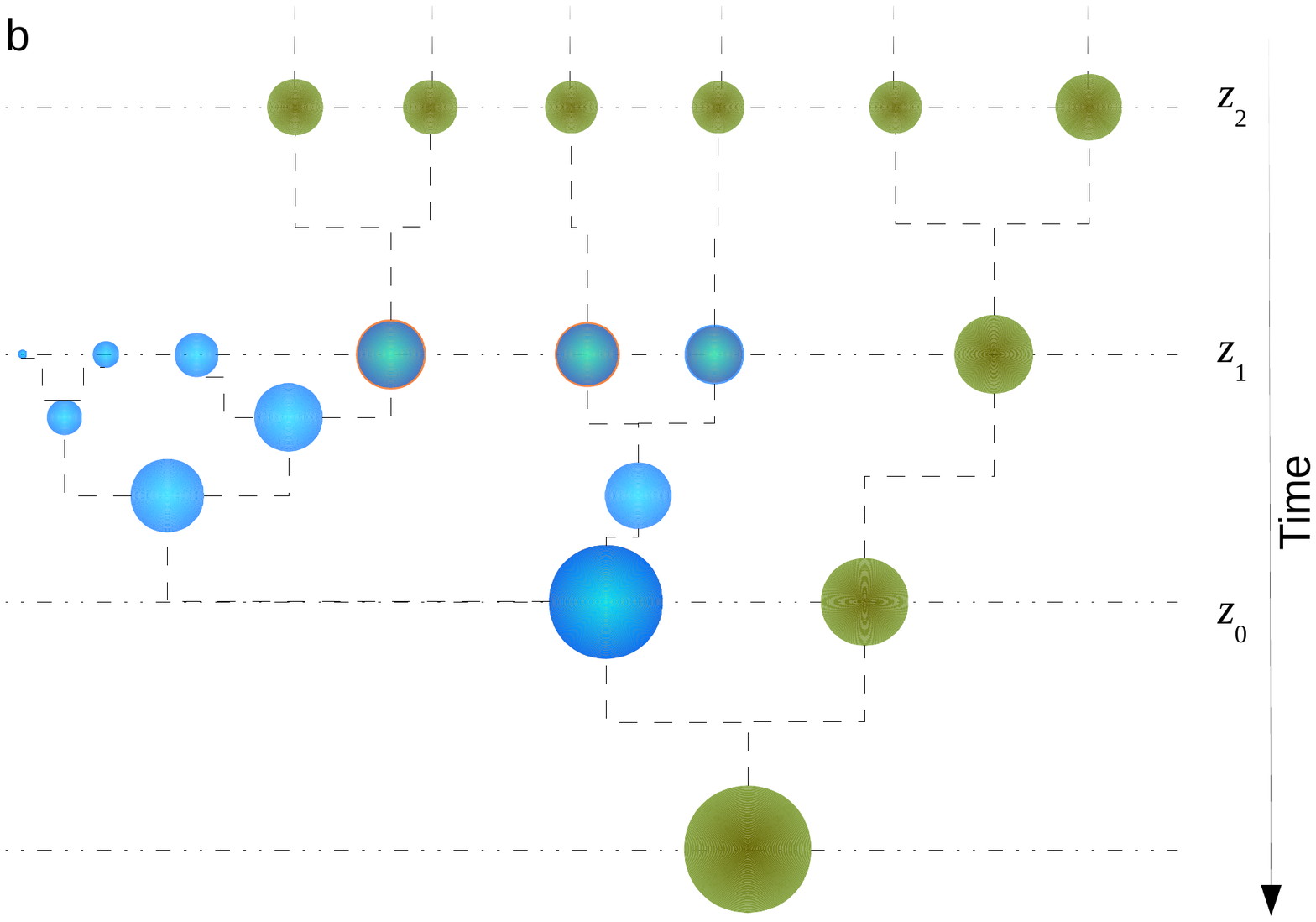} \\
   \includegraphics[width=75mm,bb=0 300 612 792]{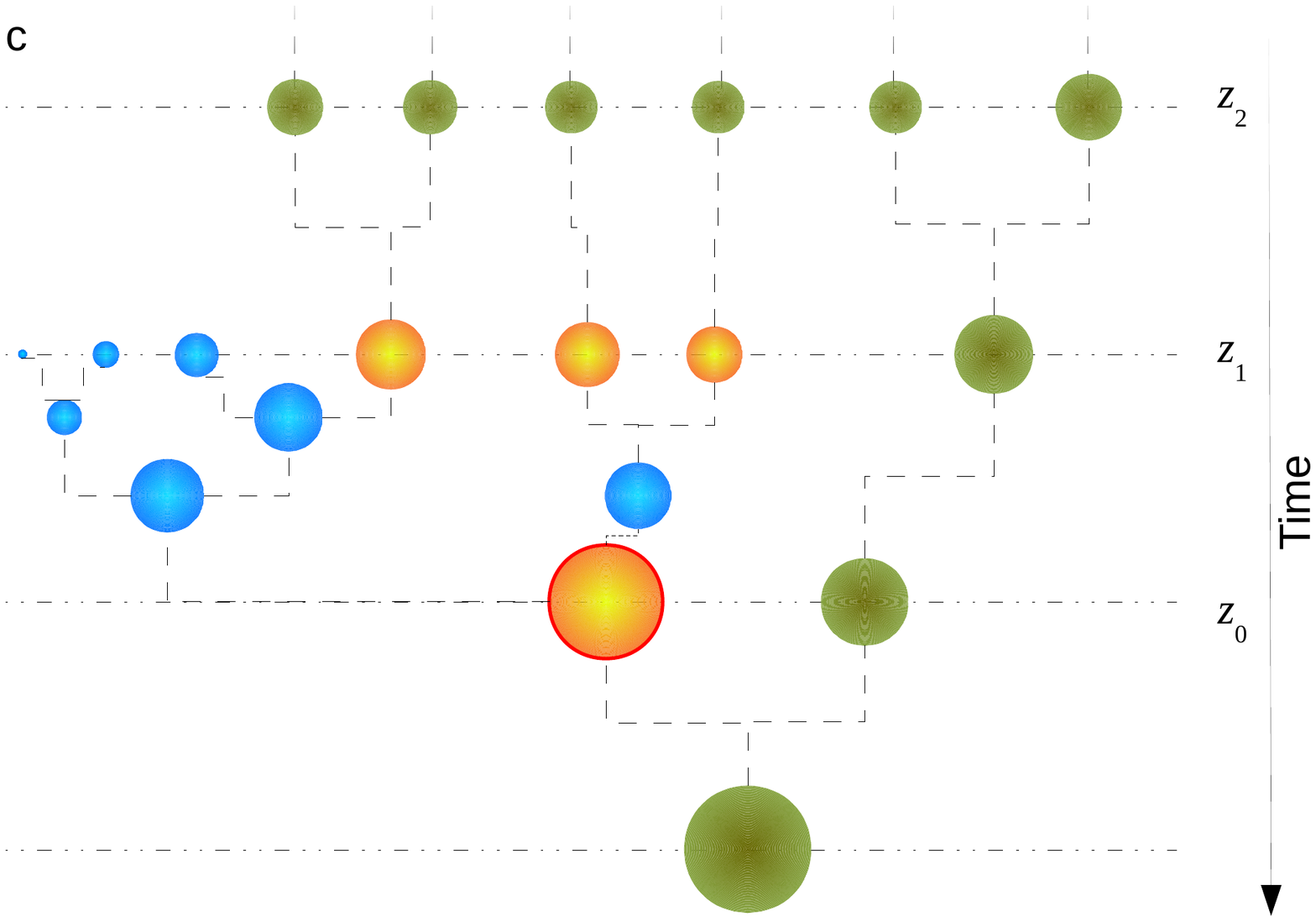} &
   \includegraphics[width=75mm,bb=0 300 612 792]{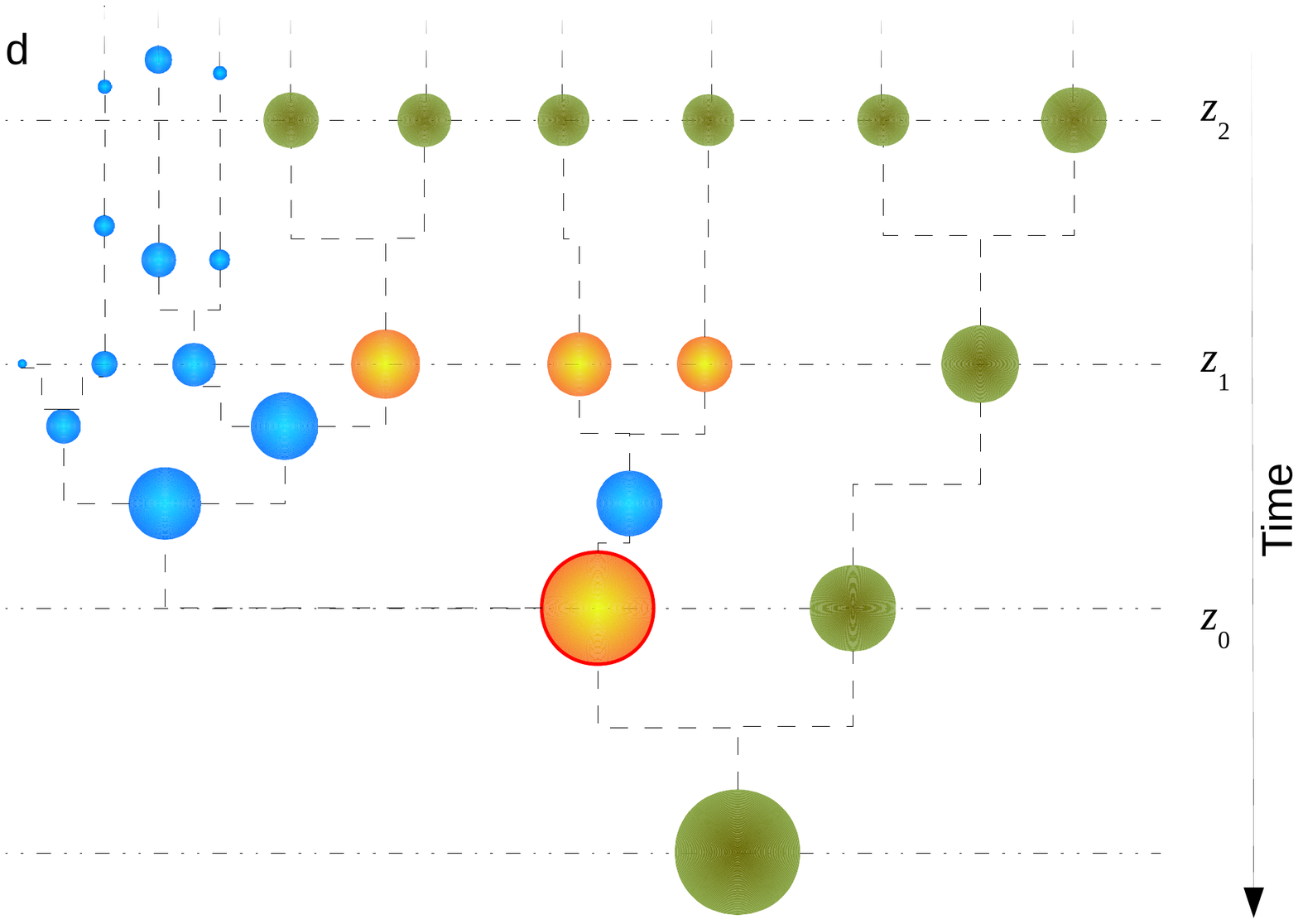}
 \end{tabular}
 \caption{A schematic showing our how tree augmenting algorithm works. \textit{Panel a:} A simplified diagram of a merger tree extracted from an N-body simulation. Circles represent halos (with radius indicating mass). Time increases in the direction shown by the arrow, and halos are located at quantized redshifts, labeled $z_0$\ldots$z_2$, and shown by horizontal, dot-dot-dashed lines. Dashed lines connect halos to their direct progenitors. One halo at $z_0$ (highlighted by the yellow color and red outline) is selected for augmentation. This halo's progenitors at $z_1$ are also highlighted in yellow. \textit{Panel b:} A trial tree (shown in blue) is generated using the PCH algorithm and compared to the halos of interest in the original tree. In this case, the match between trial tree and original tree halo masses is sufficiently close to be deemed acceptable. \textit{Panel c:} The accepted trial tree is grafted into the original tree. Note that the augmented node and its progenitors from the original tree are retained (so their masses are unchanged), but now with the structure of the trial tree grafted between them. \textit{Panel d:} Where the trial tree has halos at $z_1$ which did not match any halo in the existing tree (and are below the resolution of the original tree by construction), a new tree is grown from each such halo and attached. This process is repeated for each of the green halos in the original tree.}
 \label{fig:schematic}
\end{figure*}

We consider the N-body merger tree as a graph consisting of a collection of connected nodes (halos; see Figure~\ref{fig:schematic}a). In this work we will exclusively use halo trees, as opposed to subhalo trees. Our method is applicable to subhalo trees also, but we choose not to explore such trees in this work as explained in \S\ref{sec:GalaxyConvergence}. Each node is described by a mass and quantized epoch (i.e. a redshift taken from the set of snapshot redshifts, $z_0\ldots z_N$ (with $z_0<z_1<z_2\ldots$), output by the N-body simulation), and a collection of zero or more immediate progenitor nodes (i.e. those connected nodes existing at the immediate prior snapshot redshift). In the \cite{parkinson_generating_2008} algorithm the statistics of progenitors of a halo depend only on the mass and redshift of that halo. Therefore, with no loss of generality, for any given node we relabel redshifts such that the halo exists at $z_0$. Therefore, we label the node with the tuple $(M_0;z_0)$, and the masses and redshift of its immediate progenitors as $(M_1,M_2,\ldots,M_n;z_1)$, where $n$ is the number of progenitors and $M_1 \ge M_2 \ge M_3\ldots$. Should a halo have one or more immediate progenitors which exist at $z>z_1$---which may happen if the halo finding algorithm used to build the tree was unable to locate at $z_1$ a progenitor which \textit{was} found at $z_{>1}$---we interpolate that progenitor to $z_1$ (assuming linear growth of mass with time for the primary progenitor, and no mass growth for non-primary progenitors).

Our procedure for augmenting the tree, which is shown schematically in Figure~\ref{fig:schematic}, is as follows. We first prune branches of the merger tree which begin in halos with masses below a threshold, $M_{\rm cut}$, chosen such that above this mass the structure of the tree is reliably determined and not affected by the limited resolution of the simulation. Then, we visit each node in turn and apply the following procedure:
\begin{enumerate}
 \item Grow a merger tree to the desired mass resolution, $M_{\rm res}$, using the \cite{parkinson_generating_2008} algorithm, starting from a base halo of $(M_0;z_0)$ and stopping at $z_{\rm 1}$ (or with no stopping redshift in the case of a node at the earliest snapshot of the simulation, $z_N$);
 \item We label the masses of the progenitors at $z_{\rm 1}$ in this trial tree $(M^\prime_1,M^\prime_2,\ldots,M^\prime_{n^\prime})$, with $M^\prime_1 \ge M^\prime_2 \ge M^\prime_3\ldots$.
 \item Accept this trial tree as a match to the original tree structure if:
 \begin{enumerate}
  \item $n^\prime \ge n$;
  \item $|M_i^\prime-M_i| < \epsilon M_i$ for $i=1\ldots n$;
  \item If $n^\prime > n$, $M^\prime_i < M_{\rm cut}$ for all $i=n+1\ldots n^\prime$.
 \end{enumerate}
 \item If the tree is accepted graft it in to the original tree and proceed to the next node, otherwise, return to step 1 and create a new trial tree.
\end{enumerate}

In this procedure, $\epsilon$ is a parameter which controls the mass precision required in matching the information in the original tree. Once an acceptable tree is found it is grafted into the original tree by simply replacing the base node of the new tree with the node currently being augmented in the original tree, and the progenitor nodes at $z_1$ in the new tree with the progenitors of the current node in mass-ranked order. Note that we keep the masses of the original tree nodes, we do not replace them with the masses of the matched nodes in the new tree (which will differ from those in the original tree by up to a factor of $1+\epsilon$). In this way the original information in the tree is preserved while adding in higher mass resolution branches and, potentially, a higher time resolution between redshift snapshots\footnote{This higher time resolution is achieved across all halo masses in the tree, not just in the new high resolution branches. The trial trees created in our augmenting procedure contain high time resolution structure across all halo mass scales---from the mass resolution of our augmenting procedure up to the mass of the halo being augmented. When an accepted trial tree is grafted into the original tree the full structure is retained, including halos above the resolution limit of the original simulation but which exist in between snapshots of that simulation.} (which can also be important for convergence in galaxy formation models; \citealt{benson_convergence_2012}). 

After grafting in the new branches, if $n^\prime > n$ we visit each unmatched progenitor node in the new tree which, by definition, has $M_{\rm res} < M_i < M_{\rm cut}$, and grow a new tree based at $(M_i;z_1)$ with no limit on redshift. This tree is then attached to its progenitor node.

We have implemented this algorithm in the \glc\ galaxy formation code \citep{benson_galacticus:_2011,benson_galacticus:_2012} as an operator acting on merger trees which can optionally be applied prior to galaxy formation calculations are performed, and will utilize this code to test the convergence properties of the statistics of merger trees augmented by our algorithm. We also utilize the \glc\ code\footnote{Specifically for this work we used \glc\ revision {\tt e7e891a6b00c740322d3131c31af818ad1e8686e} which can be obtained from the \glc\ repository at \href{https://bitbucket.org/abensonca/galacticus}{\tt https://bitbucket.org/abensonca/galacticus}.} to predict properties of galaxies which form in augmented and unaugmented merger trees in order to explore the how well convergence in galaxy properties is achieved using our augmenting procedure.

\subsection{Application to Millennium Simulation Trees}

To demonstrate our method, and the convergence in galaxy properties achieved, we apply this algorithm to merger trees extracted from the Millennium simulation \citep{springel_simulations_2005}. Specifically, we extract merger trees from the {\tt MPAHaloTrees..MR} table of the Millennium database\footnote{\protect\href{http://gavo.mpa-garching.mpg.de/MyMillennium}{\tt http://gavo.mpa-garching.mpg.de/MyMillennium}} \citep{lemson_halo_2006}. We choose $M_{\rm cut}=7.08\times10^{10}{\rm M}_\odot$, corresponding to 60 particles. We find that using a lower $M_{\rm cut}$ results in incorrect progenitor mass functions in the augmented trees, suggesting that these masses of lower mass halos are insufficiently reliable for our purposes. We choose an initial value for $\epsilon = \epsilon_0 \equiv 0.15$ by default---we explore in \S\ref{sec:NumericalConvergence} how sensitive the results are to the choice of this parameter. If a matched tree is not found after $N_{\rm t}=50$ trials, we increase $\epsilon \rightarrow \epsilon ( 1 + \epsilon_0 )$ and continue until a match is found\footnote{As trial trees are generated, we keep a copy of the trial tree with the best match to the original tree masses found so far. If, after increasing $\epsilon$ this tree becomes a sufficiently good match we use it instead of generating any further trial trees.}. The speed of our algorithm for a given level of convergence will be determined by the interplay of $\epsilon_0$ and $N_{\rm t}$. We explore how often $\epsilon$ must be increased to find a match in \S\ref{sec:NumericalConvergence}. We augment these trees to a variety of mass resolutions.

Additionally, we also run our galaxy formation model on the Millennium-II \citep{boylan-kolchin_resolving_2009} merger trees (specifically those from the\newline {\tt MPAHaloTrees..MRII} table). We prune these trees back to $M_{\rm cut}=5.65 \times 10^8{\rm M}_\odot$ (corresponding to 60 particles), but do not augment them---using them instead as a reference sample of higher resolution than the Millennium trees. We then augment the Millennium trees to the same resolution of $5.65\times 10^8{\rm M}_\odot$ so that they can be compared directly with the Millennium-II trees.

\section{Results}\label{sec:Results}

To demonstrate our method, we first examine convergence properties in dark matter halo conditional mass functions (\S\ref{sec:NumericalConvergence}). We then explore convergence in galaxy properties as the augmented mass resolution is increased (\S\ref{sec:GalaxyConvergence}). Finally, we augment the Millennium simulation to the resolution of the Mill\-ennium-II simulation and examine whether consistent results are obtained (\S\ref{sec:Millennium}).

\subsection{Numerical Convergence}\label{sec:NumericalConvergence}

\begin{figure}
 \includegraphics[width=76mm]{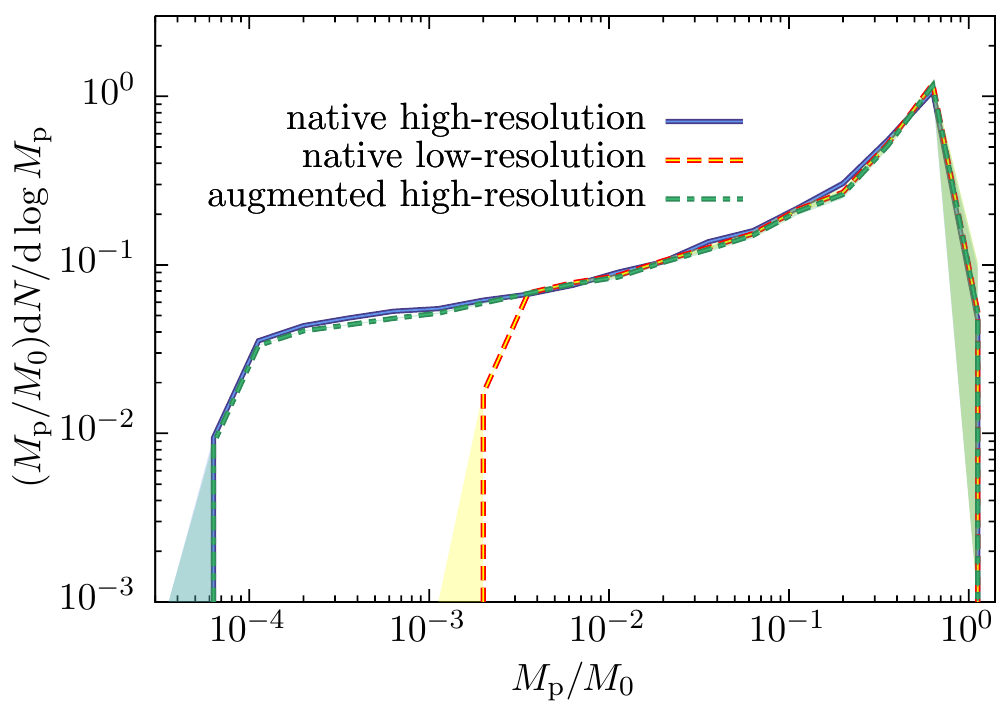}
 \caption{The conditional mass function at $z=1.08$ of halos in PCH trees which merge to become part of halos of mass $\log_{10}(M_{\rm p}/{\rm M}_\odot)=13.42$--$13.66$ by $z=0$, as a function of the progenitor halo mass normalized to the $z=0$ halo mass. The blue line indicates trees grown directly to a resolution of $M_{\rm res}=2.5\times10^9{\rm M}_\odot$ (with the shaded region indicating the $1$-$\sigma$ Poisson error bars due to the finite number of trees analyzed). The red line shows results for trees grown to a lower resolution of $M_{\rm res}=7.1\times10^{10}{\rm M}_\odot$. Finally, the green line shows results when these lower resolution trees are augmented to $M_{\rm res}=2.5\times10^9{\rm M}_\odot$. The augmenting tolerance parameter was set to $\epsilon_0=0.15$.}
 \label{fig:testPCH}
\end{figure}

Since our algorithm utilizes PCH trees to perform augmenting, a simple test of whether our algorithm works as expected is to compare the statistics of PCH trees grown directly to some high resolution with those of PCH trees grown initially to some lower resolution and then augmented to the higher resolution. Figure~\ref{fig:testPCH} shows an example of such a test, and clearly shows that the conditional mass function of augmented trees agrees very closely with that of trees grown directly to the same resolution.

Our method has two adjustable, numerical parameters, $\epsilon_0$ and $N_{\rm t}$, which control the precision required in matching halo masses between the original and proposed trees and the number of trial trees generated between successive reductions in the match precision required. To explore the sensitivity of our method to $\epsilon_0$ we apply it to merger trees from a subset of the Millennium simulation. Trees are pruned as described above, then augmented to a resolution of $2.5\times 10^9{\rm M}_\odot$. We repeat this process for $\epsilon_0=0.1500$, $0.0750$, and $0.0375$, and construct conditional mass functions of the resulting merger trees.

\begin{figure*}
 \begin{tabular}{cc}
 \includegraphics[width=76mm]{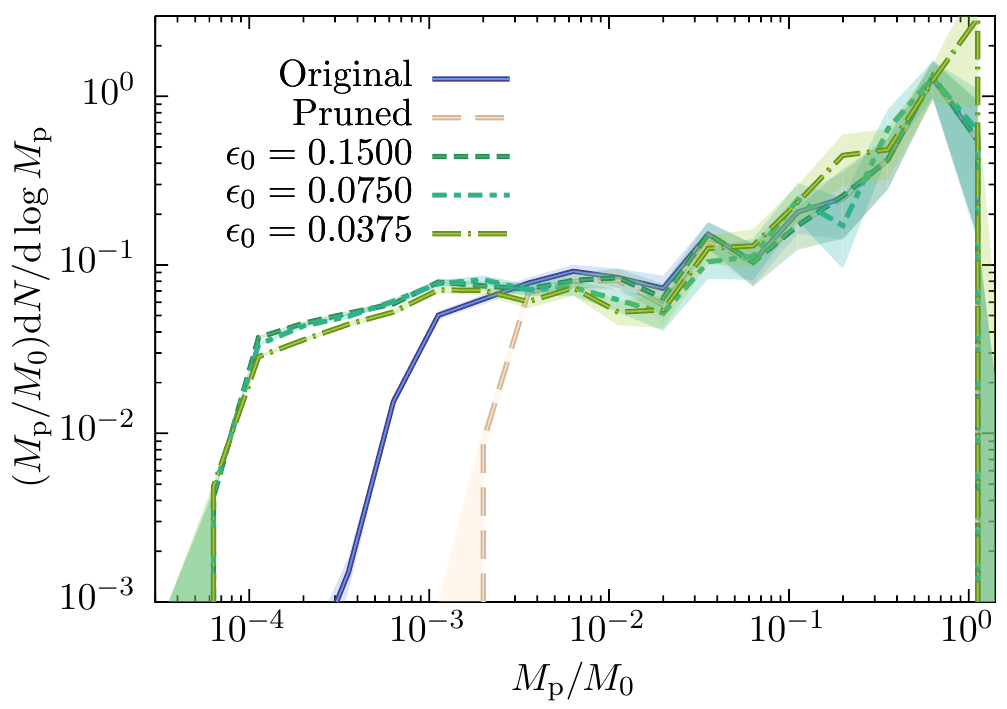} &
 \includegraphics[width=76mm]{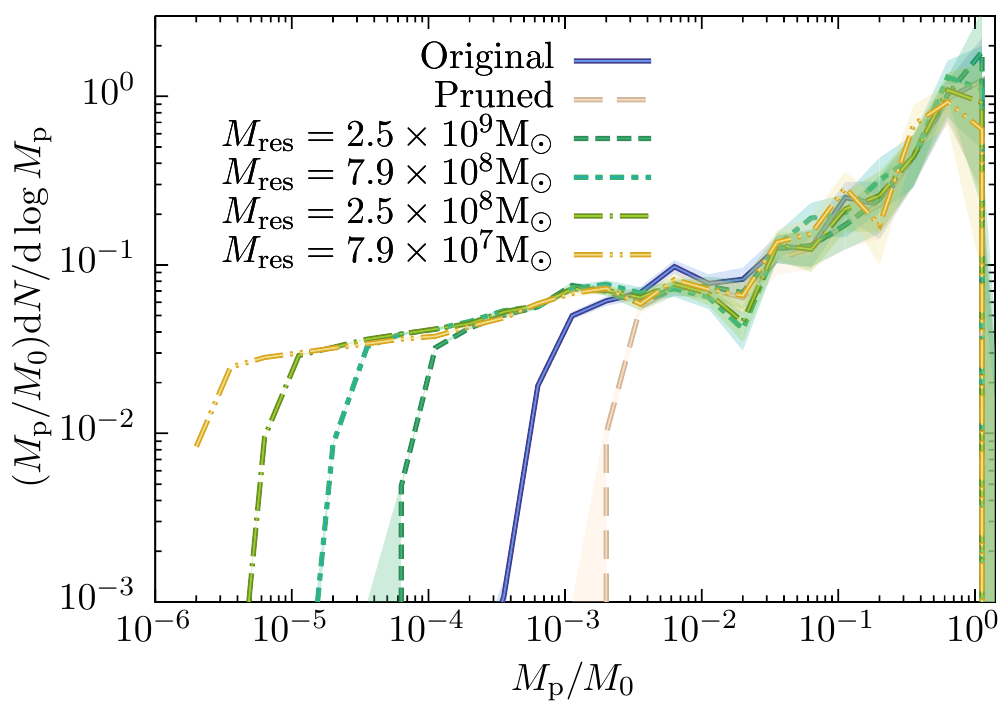}
 \end{tabular}
 \caption{\textit{Left panel:} Convergence of the conditional mass function with the $\epsilon_0$ parameter. Lines show the conditional mass function at $z=1.08$ of halos which merge to become part of halos of mass $\log_{10}(M_{\rm p}/{\rm M}_\odot)=13.42$--$13.66$ by $z=0$, as a function of the progenitor halo mass normalized to the $z=0$ halo mass. The blue line indicates the original Millennium trees (with the shaded region indicating the $1$-$\sigma$ Poisson error bars due to the finite number of trees analyzed), while the pale pink line shows results for trees after mass pruning to a resolution of $7.08\times10^{10}{\rm M}_\odot$. The three green lines indicate results for trees augmented to a resolution of $2.5\times10^9{\rm M}_\odot$ using our method for three different values of $\epsilon_0$, as indicated in the panel. \textit{Right panel:} The same conditional mass function as in the left panel, with blue and pale pink lines again showing results for the original and pruned trees. Remaining lines show results after augmenting trees to different mass resolutions as indicated in the panel. All cases use $\epsilon_0=0.15$.}
 \label{fig:convergence}
\end{figure*}

The left-panel of Figure~\ref{fig:convergence} shows convergence in the conditional mass function with the $\epsilon_0$ parameter. Specifically, we show the conditional mass function at $z=1.08$ of halos which merge to become part of halos of mass $\log_{10}(M_{\rm p}/{\rm M}_\odot)=13.42$--$13.66$ by $z=0$, as a function of the progenitor halo mass normalized to the $z=0$ halo mass. The blue line indicates the original Millennium trees (with the shaded region indicating the $1$-$\sigma$ Poisson error bars due to the finite number of trees analyzed), while the pale pink line shows results for trees after mass pruning to a resolution of $7.08\times10^{10}{\rm M}_\odot$. The three green lines indicate results for trees augmented to a resolution of $2.5\times10^9{\rm M}_\odot$ using our method for three different values of $\epsilon_0$, as indicated in the panel.

We find that 98.5\%, 96.9\%, and 92.0\% of bran\-ches were matched to within the original tolerance for $\epsilon_0=0.1500$, $0.0750$, and $0.0375$ respectively---even though we allow this tolerance to fall off if a matching tree is not found after $N_{\rm t}=50$ trials it is clear that the majority of cases are actually matched at the required initial tolerance. In particular, we find that the conditional mass function is well-converged over the full range of masses as $\epsilon_0$ is varied. This indicates that a value of $\epsilon_0=0.15$ is sufficient to ensure convergence.

Considering the parameter $N_{\rm t}$, we find that, for $\epsilon_0=0.15$, for nodes with only a single progenitor node a match is found after an average of $1.8$ trial trees. These cases are typically easy to match as (by selection) these nodes do not have any mergers above the resolution limit over the period being augmented. For nodes with two or more progenitor halos matching becomes more challenging. In those cases we find that a match is found after an average of $20$ trial trees. We find that as $\epsilon_0$ is decreased this mean number of trial trees before a match is found increases only slowly (e.g. it is $22$ for $\epsilon_0=0.0375$). If we increase $N_{\rm t}$ we find that the mean number of trial trees before a match is found also increases. For example, with $\epsilon_0=0.15$, the mean number of trial trees\footnote{These means include only cases where a successful match was found within $N_{\rm t}$ trials.} is $19.6$, $33.8$, $52.3$ for $N_{\rm t}=50$, $100$, $200$ respectively. Since each trial is independent, if the probability for any given tree to match successfully is $p$, then the probability distribution for a match after $n$ trials is simply $f(n)=p(1-p)^{n-1}$. The mean number of trials, $\langle f(n) \rangle$ is then approximately consistent with the above results for $N_{\rm t}=50$, $100$, and $200$ if $p=0.0167$. For $\epsilon_0=0.0375$ we find $p\approx 0.0067$. Based on these results, it is clear that $N_{\rm t}$ could be increased without significant loss of speed, and with some improvement in accuracy. Specifically, for $\epsilon_0=0.15$ and the specific tree resolutions considered in these tests, $N_{\rm t}=50$ will result in around 55\% of two-or-more progenitor branches being matched at the original tolerance, $\epsilon_0$, while to have 90\% of such branches matched at the original tolerance would required $N_{\rm t}\approx 140$. As we will show below, $N_{\rm t}=50$ is sufficient to achieve good convergence in all tests that we consider.

The right-panel of Figure~\ref{fig:convergence} shows the same conditional mass function, but now explores convergence as the augmented mass resolution is changed at fixed $\epsilon_0=0.15$. As mass resolution in the augmented branches is increased the conditional mass function clearly converges at fixed mass ratio (of course, at higher resolution the conditional mass function is populated down to lower mass ratio). This demonstrates that our method is stable as mass resolution is changed.

\subsection{Galaxy Property Convergence}\label{sec:GalaxyConvergence}

To explore convergence in galaxy properties we examine the stellar mass vs. halo mass relation. As mentioned in \S\ref{sec:Method} we use halo trees in this work, as opposed to subhalo trees. The \glc\ model used in this work is configured to not require any information from N-body subhalos---merging times for subhalos (whether these were part of the original trees, or high resolution subhalos grafted into those trees) are computed using the fitting function of \cite{jiang_fitting_2008} with orbital parameters drawn from the distribution of \cite{benson_orbital_2005}. Our augmenting method will also work with subhalo trees. However, if N-body subhalo information (such as positions, orbital parameters, etc.) from such trees were used by a semi-analytic model, then equivalent information for the high resolution grafted subhalos would have to be treated via (semi-)analytic methods. Such an approach will be the subject of a future work.

\begin{figure}
 \includegraphics[width=75mm]{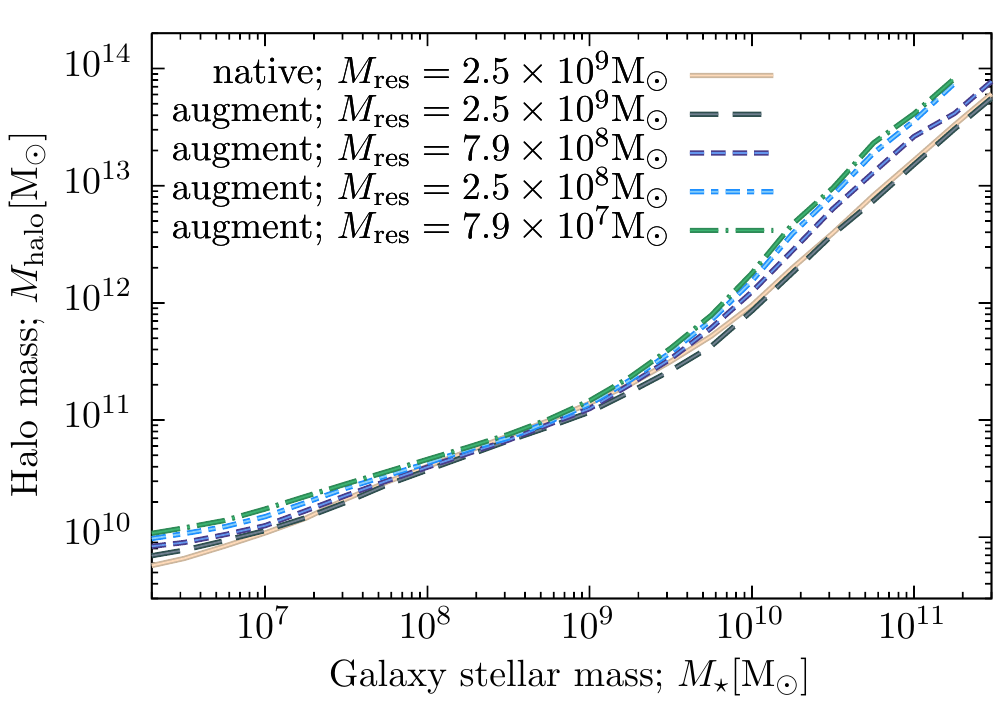}
 \caption{The stellar mass vs. halo mass relation as a function of augmented tree mass resolution. The ``native'' line shows results for trees built to a resolution of $2.5\times10^9{\rm M}_\odot$ entirely using the PCH algorithm. Other lines show results for Millennium simulation merger trees augmented to different mass resolutions as indicated in the figure.}
 \label{fig:convergenceGalaxies}
\end{figure}

Figure~\ref{fig:convergenceGalaxies} shows the stellar mass vs. halo mass relation as computed by the \glc\ model. Note that we are not interested in whether this relation (which depends on the specific implementation of baryonic physics in our model) agrees well with observations, but merely whether it converges as we augment trees to higher resolution. Of course, the details of convergence will depend on the details of the galaxy formation physics modeled. However, providing that galaxy formation becomes inefficient in some sufficiently low mass halos convergence should always be reachable. The ``native'' lines shows results for merger trees built entirely via the PCH algorithm, with $M_{\rm res}=2.5\times10^9{\rm M}_\odot$. Other lines show results from Millennium merger trees augmented to various different resolutions as indicated in the figure.

When augmented to a mass resolution of $2.5\times10^9{\rm M}_\odot$ the Millennium trees produce results in good agreement with the ``native'' trees. As the resolution is increased there is a clear systematic shift in the stellar mass vs. halo mass relation---with the halo mass at fixed stellar mass increasing at both low and high masses, while remaining almost constant around the bend in the relation at a few times $10^9{\rm M}_\odot$ in stellar mass. Importantly though, it is clear that the relation is converging as the tree resolution is increased. This is expected---in this particular model infall of gas into halos from the intergalactic medium is inhibited in halos with circular velocities at their virial radius below 35~km~s$^{-1}$ (corresponding to halo masses of $2.05\times10^{10}{\rm M}_\odot$ at $z=0$). Once the tree resolution is high enough to resolve all such halos we expect convergence in the results.

\subsection{Millennium vs. Millennium-II}\label{sec:Millennium}

As a further demonstration of our method, we use it to augment merger trees from the Millennium simulation (mass resolution $7.08\times 10^{10}{\rm M}_\odot$ after pruning) to match the mass resolution of the Millennium-II simulation ($5.65\times10^8{\rm M}_\odot$ after pruning). We augment roughly 3.5\% of the volume of the Millennium simulation. Since the Millennium-II simulation has a total volume 125 times smaller than the Millennium simulation this means our augmented 3.5\% of the Millennium volume corresponding to roughly 439\%(=$125\times 3.5\%$) of the Millennium-II simulation volume. We then run our galaxy formation model on these augmented trees to predict galaxy properties.

\begin{figure*}
 \begin{tabular}{cc}
 \includegraphics[width=75mm]{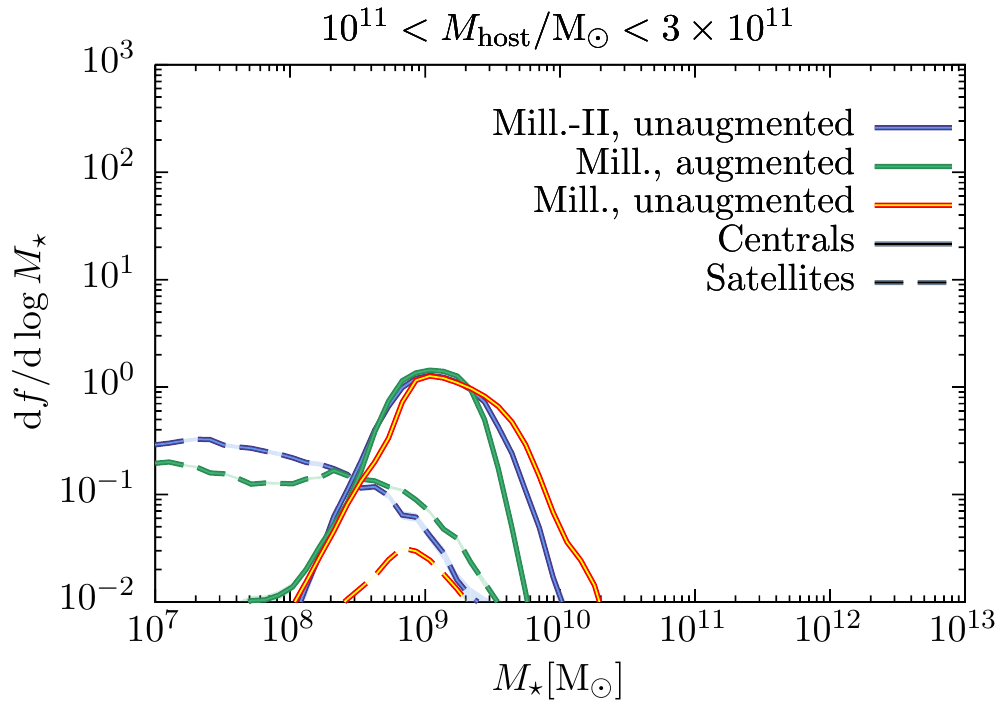} &
 \includegraphics[width=75mm]{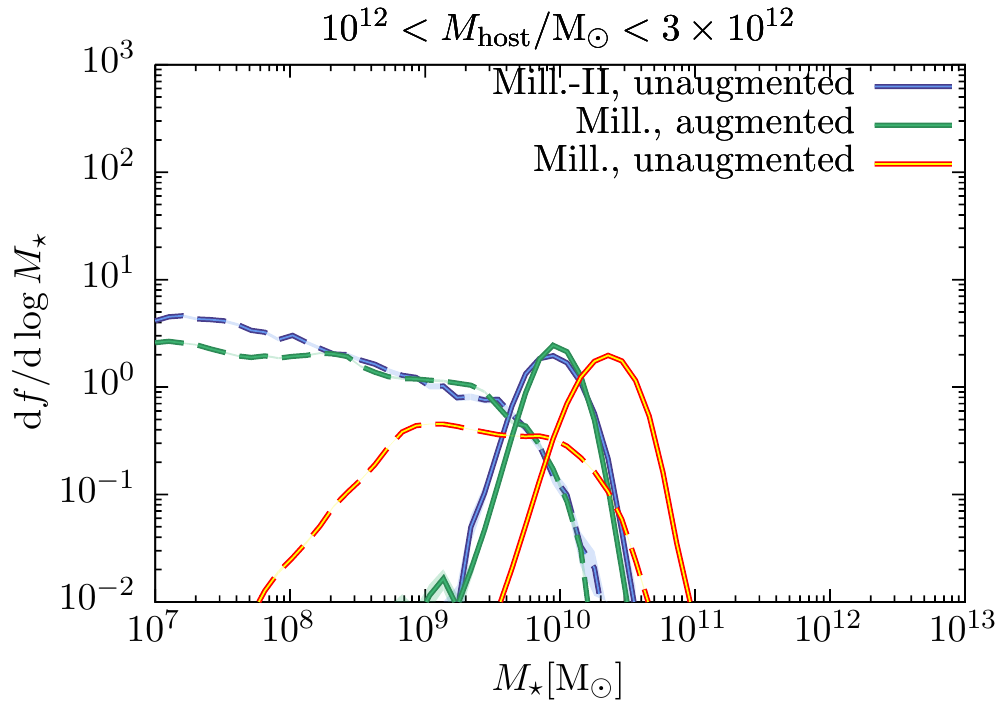} \\
 \includegraphics[width=75mm]{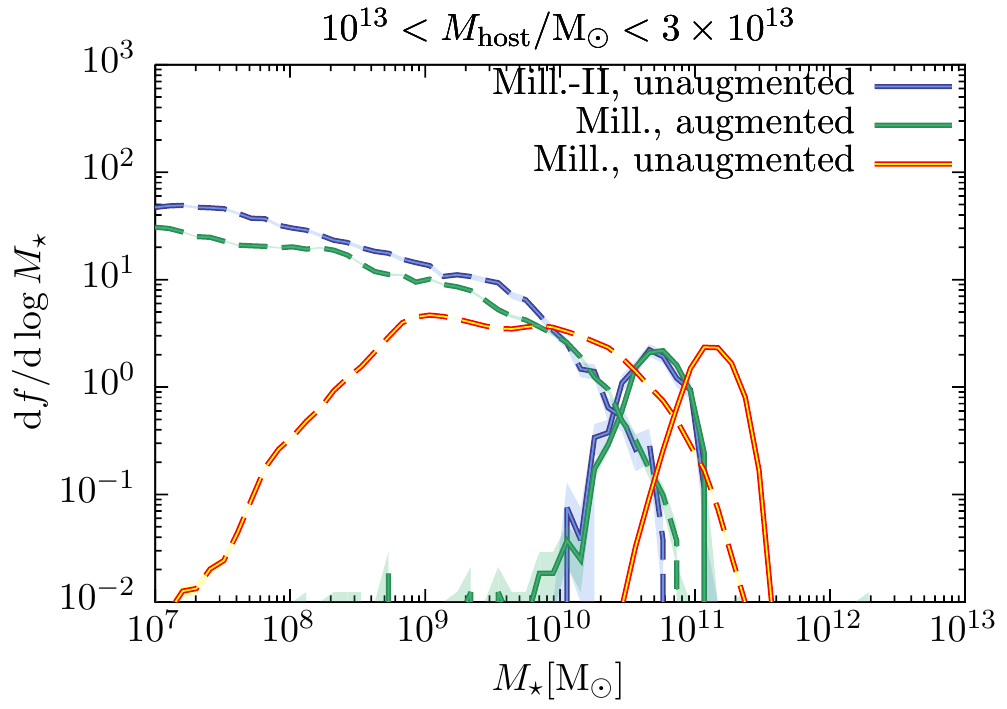} &
 \includegraphics[width=75mm]{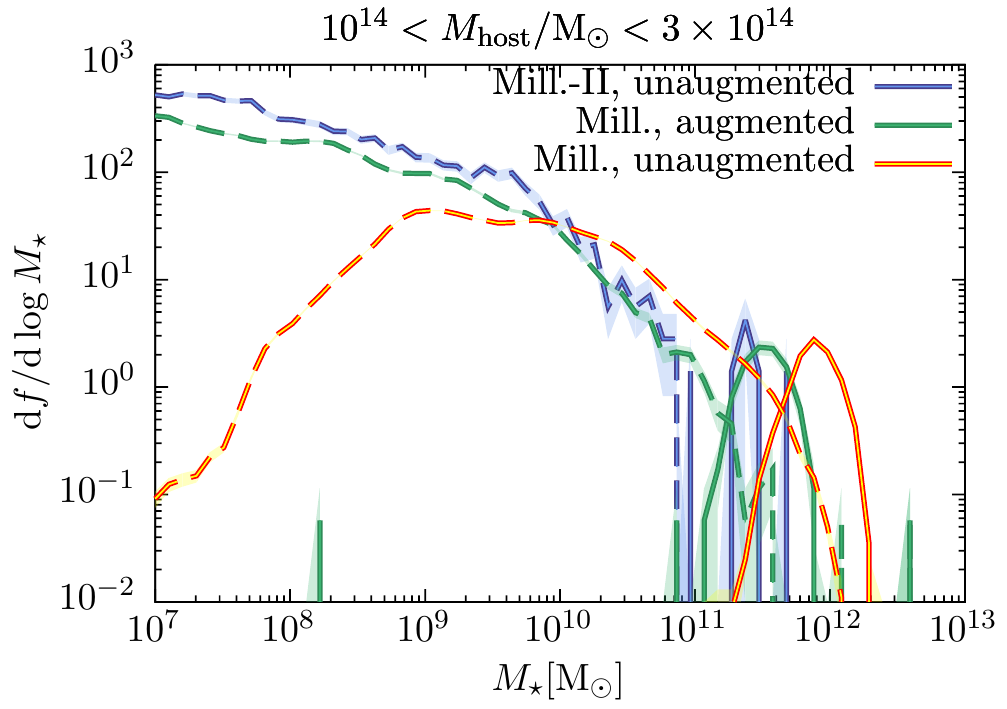} 
 \end{tabular}
 \caption{Stellar mass functions of galaxies in the Millennium and Millennium-II simulations. Each panel shows the distribution of galaxy stellar masses in $z=0$ halos in a narrow range of masses (as shown above each panel), normalized to show the average number of galaxies within a single $z=0$ halo. Solid lines indicate masses of central galaxies, while dashed lines indicate satellite galaxies. Red lines show results for higher resolution, unaugmented (but pruned) Millennium simulation trees, blue lines show results for unaugmented (but pruned) Millennium-II simulation trees, and green lines show results for Millennium trees augmented to match the resolution of the Millennium-II trees. Shaded bands indicate $1$-$\sigma$ Poisson errors on the mass functions arising from the finite number of merger trees analyzed.}
 \label{fig:convergenceMillennium}
\end{figure*}

In Figure~\ref{fig:convergenceMillennium} we show stellar mass functions of galaxies in the Millennium and Millennium-II simulations, normalized to show the average number of galaxies within a single $z=0$ halo. Each panel shows the distribution of galaxy stellar masses in $z=0$ halos in a narrow range of masses (as shown above each panel). Solid lines indicate masses of central galaxies, while dashed lines indicate satellite galaxies. Red lines show results for unaugmented (but pruned) Millennium simulation trees, blue lines show results for unaugmented (but pruned) Millennium-II simulation trees, and green lines show results for Millennium trees augmented to match the resolution of the Millennium-II trees. The unaugmented Millennium and Millennium-II trees show clear differences in these mass functions, which must arise due to their different resolutions. The augmented Millennium trees produce results clearly much closer to those from Millennium-II trees, although some differences remains. Before assessing the success of the augmenting alogorithm in this case we will first examine whether the residual effects of mass resolution are plausibly affecting these results.

To assess the expected effects of resolution on this plot we identify galaxies in halos of mass between $1$ and $2$ times the resolution limit of the pruned Millennium-II trees. We then find the distribution of galaxy stellar masses in such halos and identify the $84^{\rm th}$ percentile of that distribution. We find this to be approximately at $M_\star=3.7\times 10^6{\rm M}_\odot$. As defined, this resolution limit therefore lies below the range plotted in Figure~\ref{fig:convergenceMillennium}. Furthermore, we can select galaxies in the stellar mass range $M_\star=1$--$2\times 10^7{\rm M}_\odot$ (i.e. the lowest masses shown in Figure~\ref{fig:convergenceMillennium}) and examine the distribution of halo masses. We find that the $16^{\rm th}$ percentile of this distribution is $M_{\rm halo}=8.7\times 10^9{\rm M}_\odot$---over 10 times higher than the resolution of our pruned Millennium-II trees. Nevertheless, this does not guarantee that the mass functions shown in Figure~\ref{fig:convergenceMillennium} will be completely free from resolution artifacts, since these $8.7\times 10^9{\rm M}_\odot$ halos are, of course, built from yet smaller halos closer to the resolution threshold.

Focusing first on central galaxies we see that such galaxies in unaugmented Millennium trees are biased to higher stellar mass (at fixed halo mass) than those in (unaugmented) Millennium-II trees. In this specific model, this occurs because baryons are locked up as stars and interstellar medium in lower masses galaxies in the high resolution Millennium-II trees, making them unavailable in the formation of the central galaxy. In the lower resolution Millennium trees, these baryons remain available for incorporation into the central galaxy since the low mass galaxies never form. However, after augmenting the Millennium trees to match the resolution of the Millennium-II trees we see that central galaxy stellar mass distributions are almost identical in Millennium and Millennium-II. 

For satellite galaxies we see a similar behavior---in unaugmented Millennium trees satellite galaxies tend to be biased to higher masses than their counterparts in Millennium-II trees. After augmenting the satellite mass functions agree closely at high masses in Millennium and Millennium-II, although we see some differences between the satellite mass functions, even after augmenting. This is particularly noticeable in the top-left panel for higher mass central galaxies, and is apparent in all panels for low mass satellite galaxies. This may be due to limitations of our approach, but may also simply reflect differences in how resolution effects depend on mass in N-body and PCH trees. To investigate these possibilities we repeat this study but instead of using trees extracted from the Millennium and Millennium-II simulation we generate a set of PCH trees matched to the resolutions of those two simulations (referring to these as ``pseudo-Millennium'' and ``pseudo-Millennium-II'' trees) and then attempted to augment the lower resolution trees to match the higher resolution trees. In this case, by construction, the mass dependence of the resolution cut-off in tree branches is identical between the two sets of trees. We find that the satellite mass functions agree very well between augmented pseudo-Millennium and pseudo-Millennium-II trees. As such, differences in the mass-dependence of the resolution cut-off between PCH and N-body trees seem to be the cause of the differences in the low mass regions of satellite mass functions in Figure~\ref{fig:convergenceMillennium}---in the top left panel the mass to which the Millennium trees are pruned before augmenting, $7.08\times10^{10}{\rm M}_\odot$, is very close to the range of halo masses plotted, 1--$3\times 10^{11}{\rm M}_\odot$, so it is not too surprising that resolution has some residual effects.

In any case, we expect that in typical applications of our approach the important issue will be that galaxies in halos above the resolution limit of the original simulation (corresponding to the low mass turnovers in the unaugmented Millennium mass functions) have converged properties---since it is these halos for which spatial information is available. Clearly, in this example such convergence is achieved.

\section{Conclusions}\label{sec:Conclusions}

We have described a simple yet powerful way to augment the resolution of merger trees extracted from N-body simulations of structure formation by grafting in high resolution branches generated using the PCH algorithm. These grafted branches are chosen to be consistent with the existing halo masses of the tree, while providing statistically representative structure and halo masses beyond the resolution of the original tree.

We have demonstrated that our method produces results that are converged with respect to its numerical tolerance parameter, and with the mass resolution to which trees are augmented. Additionally, by applying a semi-analytic galaxy formation model to the augmented trees we have shown that galaxy properties converge as tree resolution is increased, and that we can successfully augment trees from a low resolution simulation to match the resolution of a higher resolution simulation and that the resulting galaxy properties are in excellent agreement between the two cases.

The approach of applying Markov Chain Monte Carlo techniques to semi-analytic models of galaxy formation in order to constrain their parameters and match the statistical properties of the observed galaxy population \citep{henriques_monte_2009,bower_parameter_2010,lu_bayesian_2011,mutch_constraining_2013,benson_building_2014,ruiz_calibration_2015} has become invaluable in allowing the construction of accurate galaxy catalogs, and for exploring the physics of galaxy formation. Such calibrations should of course be carried out using merger trees of sufficiently high resolution that the galaxy properties are converged. Unfortunately, if the constrained model is then to be applied to a cosmological N-body simulation, typically with lower resolution, the resulting galaxy properties may shift away from their constraints. Our approach solves this problem by allowing sufficiently high resolution to be attained even in low resolution cosmological simulations.

\section*{Acknowledgments}

The Millennium Simulation databases used in this paper and the web application providing online access to them were constructed as part of the activities of the German Astrophysical Virtual Observatory (GAVO). SC acknowledges support by the Science and Technology Facilities Council [ST/L00075X/1]. AJB acknowledges valuable discussion with Yu Lu. We acknowledge the support of the Ahmanson Foundation through the provision of computational resources used in this work.

\bibliographystyle{spbasic}
\bibliography{augmentAccented}

\end{document}